# Temporal Loss Boundary Engineered Photonic Cavity

Longqing Cong[1,*], Jiaguang Han[2], Weili Zhang[3], and Ranjan Singh[4]

[1]*Department of Electrical and Electronic Engineering, Southern University of Science and Technology, Shenzhen 518055, China*

[2]*Center for Terahertz Waves and College of Precision Instrument and Optoelectronics Engineering, Tianjin University, and the Key Laboratory of Optoelectronics Information and Technology (Ministry of Education), Tianjin 300072, China*

[3]*School of Electrical and Computer Engineering, Oklahoma State University, Stillwater, OK 74078, USA*

[4]*Division of Physics and Applied Physics, School of Physical and Mathematical Sciences and Centre for Disruptive Photonic Technologies, The Photonics Institute, Nanyang Technological University, Singapore 637371, Singapore*

[*]Email: conglq@sustech.edu.cn

**Abstract:** Losses are ubiquitous and unavoidable in nature inhibiting the performance of most optical processes. Manipulating losses to adjust the dissipation of photons is analogous to braking a running car that is as important as populating photons via a gain medium. Here, we introduce the transient loss boundary into a photon populated cavity that functions as a 'photon brake' and probe photon dynamics by engineering the 'brake timing' and 'brake strength'. Coupled cavity photons can be distinguished by stripping one photonic mode through controlling the loss boundary, which enables the transition from a coupled to an uncoupled state. We interpret the transient boundary as a perturbation by considering both real and imaginary parts of permittivity, and the dynamic process is modelled with a temporal two-dipole oscillator: one with the natural resonant polarization and the other with a frequency-shift polarization. The model unravels the underlying mechanism of concomitant coherent spectral oscillations and generation of tone-tuning cavity photons in the braking process. By synthesizing the temporal loss boundary into a photon populated cavity, a plethora of interesting phenomena and applications are envisioned such as the observation of quantum squeezed states, low-loss nonreciprocal waveguides and ultrafast beam scanning devices.

**Introduction**

Strong light-matter interactions are of extreme importance in optical applications ranging from quantum electrodynamics[1] to photonic memory[2] and optical information processing[3]. Single, or multi-mode resonant cavities are critical components in rendering large quality factors ($Q$) that provide an elegant way to tailor the strength of light-matter interactions[4, 5]. Recent interests have been focused on an extra dimension that is involved as an alternative of three-dimensional cavities: temporal boundary[6], that shifts the emission frequency of a microcavity and leads to nontrivial physics such as Lorentz nonreciprocity[7, 8, 9, 10] and dynamic wavefront[11] for potential on-chip and free-space applications. Deeper physical insights with additional degree of freedom emerge when the photon-populated cavity has a long-lasting lifetime as well as long cycle period so that the injection of temporal boundary can be controlled in a sub-cycle scheme[1, 12]. Several works have demonstrated the *adiabatic* frequency shift for canonical all-optical modulation[2, 3, 13, 14, 15], but non-*adiabatic* process is challenging and requires the pump timescale shorter than the cycle of stored photons, i.e., subcycle control of light-matter interactions, where quantum squeezed states would be generated.[1, 16] While most recent studies concentrate on the frequency shift induced by the temporal boundary of the real part of permittivity[6, 17], the associated imaginary part always couples to the temporal dynamics subject to the Kramers-Kronig relation[18]. The physics becomes nontrivial when the dynamics of imaginary part (absorption) is also considered where $Q$ factor is spoiled due to perturbations[16]. However such effects are usually ignored in most previous studies[3, 6, 17]. Although absorption loss is generally adverse in most optical processes, it can be utilized as a 'knob' to tune the cavity photon dynamics that is analogous to an essential 'brake' of a running car. The long photon lifetimes of ultrahigh-$Q$ cavities are clearly attractive, but it is crucial to change cavity $Q$ dynamically with a freely controllable tradeoff between bandwidth and photon storage time, which is especially important in optical communications.

Here, we focus on transient losses and introduce the temporal loss boundary as a 'photon brake' to control the lifetime of photons in a metamaterial cavity. The temporal dimension extends our capability to distinguish coupled photons by stripping one mode from the spatiospectrally inseparable modes via selective photon damping in the time domain. The loss boundary induces transition of an intrinsically coupled cavity to an uncoupled one with much less annihilation of populated photons than conventional approaches. In the process of temporal boundary injection, we observe the emergence of coherent oscillations and tone-tuned polarization that contributes to the linear frequency tuning of cavity photons. A temporal two-dipole model nicely reproduces the experimental observations by involving both transient real and imaginary parts of the cavity.

**Results**

**Temporal and spectral interpretation.** In time domain, a single cycle (or multicycle) probe pulse couples to the cavity modes that typically radiate with oscillations at their eigenmode frequencies and damp gradually depending on their individual $Q$s (Fig. 1a). Far-field signal sampling maps the output profile with most spectral energy scattered in the main pulse, and the cavity modes manifest themselves as the long-lasting oscillations (Fig. 1b). A temporal loss boundary is introduced at a certain instant that acts as the 'brake' to selectively hasten the damping of photon oscillations so that the intrinsic $Q$ of a specific cavity mode is tuned. The boundary is characterized by two freely controlled parameters: *height* and *delay* (Fig. 1c). Height determines the damping rate associated with density of injected photocarriers, and delay refers to the time difference between the onset time of signal ($t_0$) and pump-on instant ($t_p$) defined as $t_r = t_p - t_0$ (see Methods for experimental implementation). Three temporal regimes are classified according to the delay: I. $t_r \leq 0$; II. $0 < t_r < t_m$; III. $t_r \geq t_m$, where $t_m$ refers to the intrinsic oscillation lifetime of cavity photons. Effects induced by tuning the barrier height will be mainly discussed in regime I which has been studied as a conventional approach with uniform cavity modulation in a plethora of works.[13, 19, 20] The temporal boundary effects in regime II show many promising features to be explored in this work. Note that only variable of delay is adjusted with height fixed in regime II. In regime III, there is no dynamic perturbation of cavity photons captured by the probe pulse, it is therefore not discussed in this work.

In frequency domain, on the other hand, cavity is assumed to support two modes with different $Q$s in the probed frequency range at the steady state. Cavity modes are modulated by the temporal boundary in both regime I and regime II as shown in Fig. 1d (FDTD simulations, see methods), but with distinct consequences: both modes are modulated in regime I while only one mode is modulated in regime II. Note that the only variable in regime I is pump intensity (height of temporal boundary, characterized by *conductivity* of cavity material) and the only variable in regime II is pump delay ($t_r$). The underlying mechanism is revealed by tuning the variables in the two regimes, respectively. Transmission amplitude difference ($\Delta T = T_0 - T_p$, where $T_p$ and $T_0$ are transmission amplitudes with perturbation and at steady state, respectively) and $Q$ factors are shown in Fig. 1e and 1f by gradually tuning the variables. Synchronized modulation occurs for both modes in regime I, while asynchronized modulation is observed in regime II with a plateau of $Q$ (unaffected) for transverse magnetic (TM) mode. Therefore, the temporal boundary in regime II is capable to selectively control properties of a certain mode of a multimode cavity that is inaccessible in regime I.

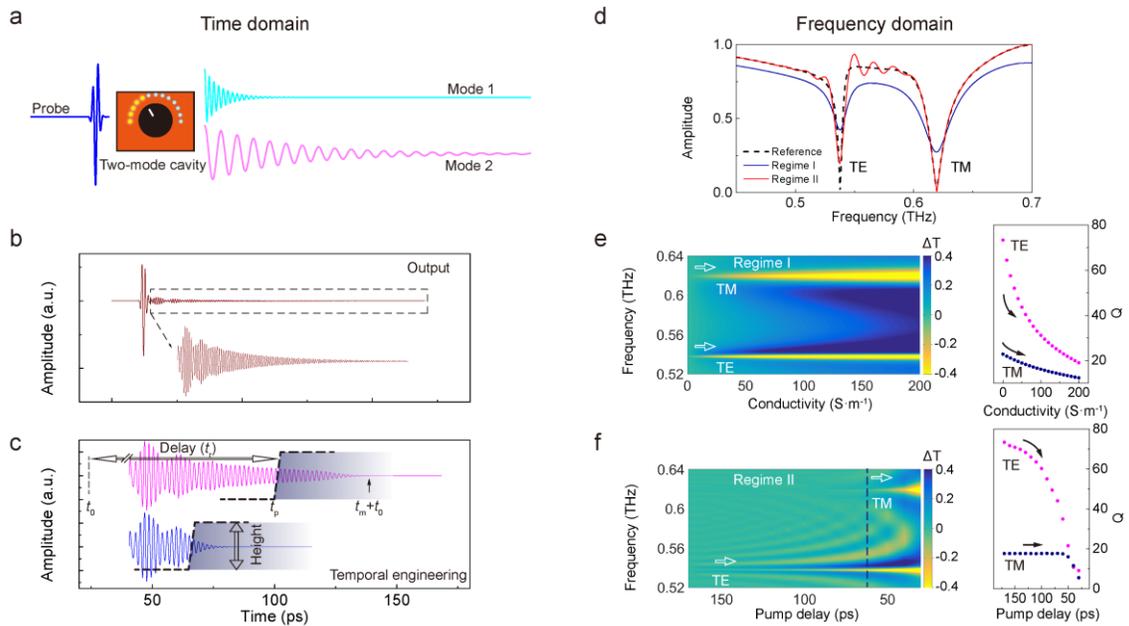

**Figure 1. Concept of temporal loss boundary interpreted from time and frequency domains.**
**a.** A two-mode cavity interpreted in time domain whose optical properties can be continuously tuned by the knob. **b.** The resultant temporal profile radiated from the two-mode cavity. **c.** The idea to tune cavity properties by introducing a temporal loss boundary that acts as a 'brake' to hasten the damping rate of oscillation at a selected instant. The two pulses indicate the same truncated oscillations engineered by loss boundary at different instants. Two parameters are defined for the temporal boundary: *delay* and *height*. Delay refers to relative time $t_r$ between the onset of the detected oscillations ($t_0$) and pump-on instant ($t_p$) of the boundary, and height indicates the strength that determines the rate of 'brake'. **d.** Modulation of the two-mode cavity interpreted in frequency domain. Dashed line shows the two modes (TE and TM) of the steady cavity; blue line shows the synchronized modulation of both modes in regime I ($t_r \leq 0$); and red line shows the asynchronized modulation spectrum in regime II ($t_r > 0$). **e** and **f**. Height- and delay-resolved spectra by tuning the boundary strength (characterized by conductivity of cavity material) and delay (characterized by time $t_r$), and the resultant evolution of $Q$ factors.

**Experimental demonstration in terahertz spectroscopy.** Terahertz time-domain spectroscopy (THz-TDS) provides an excellent platform to directly investigate the dynamics of temporal boundary, and the mechanism has potential to develop devices benefiting the next-generation wireless communications for ultrafast signal processing in this key electromagnetic band[21, 22]. The THz system maps the electric field amplitude in time domain via electro-optic (EO) sampling so that we can directly observe the temporal profiles of the cavity modes. In addition, our system covers a frequency range of 0.1 – 1.4 THz whose oscillation period lies in the picosecond range (0.7 - 10 ps) that is appropriate for temporal boundary injection as well as complete characterization of the cavity dynamics (Fig. 2a).

As a proof of concept, we designed the cavity using a metamaterial with constituents of silicon cylindrical pillars (Fig. 2b, also see ref.[20] for fabrication detail). The two modes are calibrated to be degenerate so that they couple and spatiospectrally inseparable. Experiments were carried out with the probe terahertz pulse as shown in Fig. 2a and all the frequency components lie within ~5 ps as indicated by Gabor transform (a two-dimensional Fourier transform with time and frequency dimensions, see Methods). After transmitting through the metamaterial, most frequency components are scattered within the first 10 ps through the main pulse of the output profile, and energy stored in the cavity re-emits upon scattering that prolongs the ringing for ~ 100 ps with a gradually damped amplitude (Fig. 2c). The frequency distribution of the temporal elongation is also visualized by the Gabor transform. From frequency domain, the spectrum indicates coupling of the two modes leading to a Fano lineshape as shown in Fig. 2d. The background shows corresponding eigenmode profiles with Lorentzian lineshapes, and the field distributions of a unit cell in the colored patterns indicate the transverse electric (TE) and transverse magnetic (TM) mode features with *z*-component magnetic and electric field distributions in the *z*-cut plane, respectively.

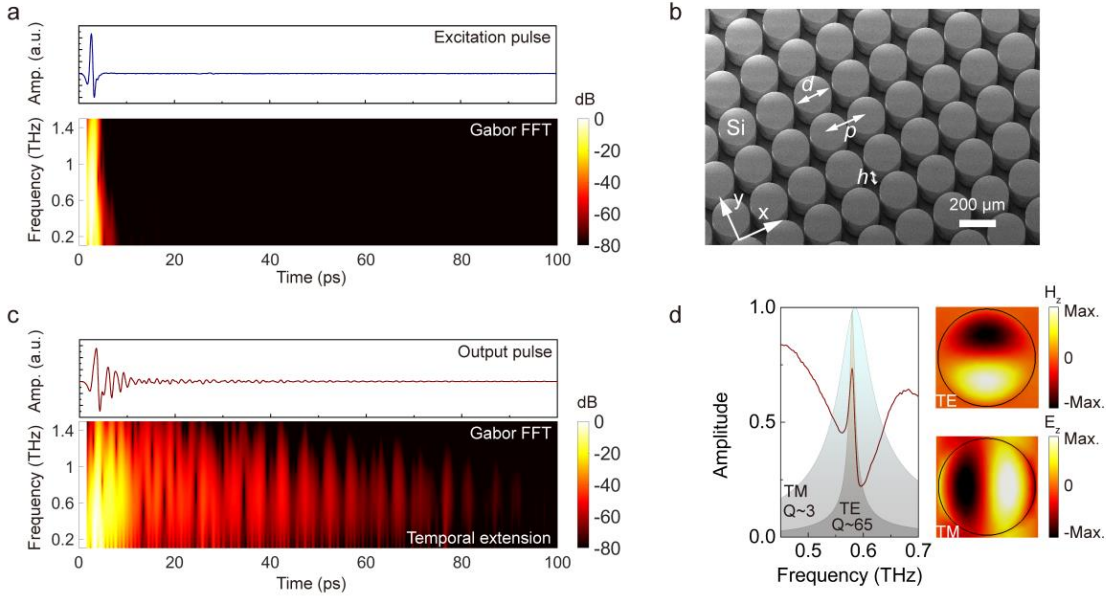

**Figure 2. Experiments with terahertz spectroscopy. a.** Temporal profile of probe pulse and its Gabor transform profile indicating the concentration of frequency components within ~5 ps. **b.** SEM image of the metamaterial whose constituents are periodic silicon cylindrical resonators with geometric parameters $d = 206$ μm, $p = 225$ μm, and $h = 120$ μm. **c.** Output temporal profile after transmitting through the cavity showing the long-lasting oscillations for ~ 100 ps. The Gabor transform reveals the elongated frequency components centered at ~ 0.6 THz. **d.** The frequency spectrum of the cavity after Fourier transform containing degenerate TE and TM modes that are spatiospectrally inseparable. Mode profiles indicate the transverse electric and transverse magnetic features in the *z*-cut plane.

An optical pump THz probe setup (OPTP, see Methods) was used to characterize the dynamics of photon-populated metamaterial which can be fully resolved by injecting free carriers on timescale much shorter (14 ps, Fig. 3b) than the cavity photon lifetimes. The pump consists of a 100 fs pulse centered at 800 nm above the bandgap of silicon to inject photocarriers whose permittivity is well described by the Drude model

$$\tilde{\varepsilon}(\omega,t) = \varepsilon'(\omega,t) + i\varepsilon''(\omega,t) = \varepsilon_\infty - \frac{\omega_{p,t}^2}{\omega^2 + i\omega\gamma} \quad , \tag{1}$$

where $\varepsilon'(\omega,t)$ and $\varepsilon''(\omega,t)$ are frequency and time dependent real and imaginary parts of permittivity, $\varepsilon_\infty$ is the high-frequency permittivity of silicon, and $\gamma$ is the electron scattering rate. The time dependent plasma frequency of photodoped silicon $\omega_{p,t}$ is related to the carrier density $N(t)$ and effective electron mass $m^*$:

$$\omega_{p,t} = \sqrt{\frac{N(t)e^2}{\varepsilon_0 m^*}}, \qquad (2)$$

where $\varepsilon_0$ is vacuum permittivity, and $e$ is the electron charge. The pump pulse generates free carriers in the conduction band and induces a sudden change in the carrier density. The relative time ($t_r$) of the pump and EO sampling pulse has a sub-picosecond resolution controlled by an optical delay line, and density of free carriers (i.e., boundary height) was adjusted by pump fluences. Photocarriers of the intrinsic silicon reveal a nanosecond relaxation which is two orders longer than rising dynamics (Fig. 3b and see Methods) so that the temporal boundary could be simply modelled as a Heaviside function.

The photocarrier induced transient imaginary part of permittivity will dominate the cavity photon dynamics. As observed in regime II, after the probe pulse has injected terahertz photons into the cavity, cavity photons are suddenly perturbed with pump energy injected. As a result, we observe the modulation of the transmission spectra associated with the pump instant (Fig. 3a). The timing of brake determines the portion of cavity photons that radiate naturally and damp artificially. In addition, the sudden merging of pump causes the emitted radiation no longer a simple superposition of Lorentzian (dashed line in Fig. 3a, and also see supplementary materials), but a perturbed mode (TE mode) involving a series of coherent oscillations along with a linewidth broadening. The high $Q$ TE mode is finally eliminated from the cavity (uncoupled cavity, will discuss later) at $t_r = 4.0$ ps with the low $Q$ TM mode almost unaffected. The long-lasting $Q$ plateau of TM mode is observed from Fig. 3d, and the modulation does not start until ~5 ps consistent with its $Q$ of ~3 while the TE mode modulation starts from ~ 100 ps (with $Q$ ~ 65). As a comparison, when the temporal boundary is set in regime I with various pump fluences, both modes reveal synchronized modulation, and the resonant radiation is always a superposition of Lorentzian as shown in Fig. 3c (dashed line and see supplementary materials) and Fig. 3e. Note that the pump is set with $t_r \ll 0$ in regime I in order to exclude the effects of photocarrier rising dynamics from interfering the cavity photons.

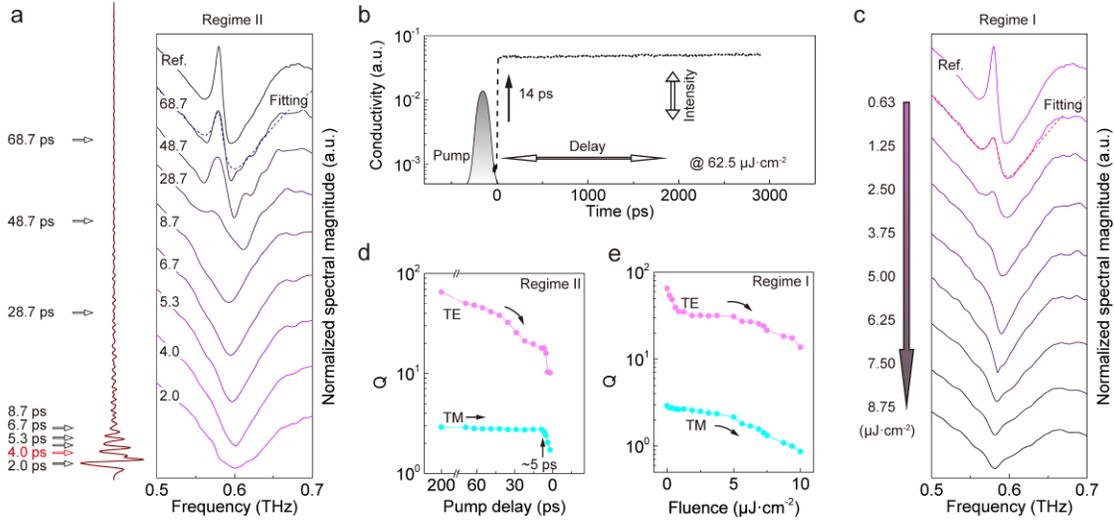

**Figure 3. Experimental verification of temporal loss boundary. a.** Spectra of perturbed cavity with injection of loss boundary at different temporal instants as indicated by the side inset. Dashed line shows the Lorentzian fitting, and pump fluence is fixed at 62.5 μJ·cm$^{-2}$. **b.** Carrier dynamics of a pristine silicon film measured by the variation of peak terahertz electric field amplitude (proportional to conductivity) due to perturbation of the ultrashort pump pulse (not scaled) with width of 100 fs. The graph shows dynamics at a pump fluence of 62.5 μJ·cm$^{-2}$, and photocarriers reach the quasi-steady state within 14 ps which lasts for longer than 3 ns. a.u.: arbitrary units. **c.** Spectra of the quasi-steady cavity at varied barrier heights in regime I, and dashed line shows the Lorentzian fitting. Here the pump is injected at $t_r \ll 0$. **d** and **e.** Modulation of $Q$ for TE and TM modes in the *delay* (d) and *height* (e) scenarios. Asynchronized modulation of TE and TM modes is observed in regime II while synchronized modulation occurs in regime I.

It is then reasonable to look for the threshold where the TE mode is decoupled from TM mode in regime II. Since the two modes are degenerate, they couple and manifest a 360° phase variation contributed by the two modes as shown in Fig. 4a. When the TE mode is completely damped, the remaining single dipole can only provide a 180° phase variation, which indicates a threshold between coupled and uncoupled states of the cavities.[23, 24] The coupling transition is visualized from the polar plot exhibiting the two coupling regimes by observing whether the spectrum circles the center point. The decoupling threshold of the specific cavity is estimated to be ~ 4 ps in regime II. Note that the coupling can also be switched in regime I at a specific pump fluence (5.0 μJ·cm$^{-2}$) when the more sensitive TE mode is completely damped out but TM mode survives (see supplementary materials). The decoupling threshold of pump fluence in regime II would be a variable corelated with $t_r$, which will finally be a constant (5.0 μJ·cm$^{-2}$) at $t_r \leq 0$ (regime I).

By comparing spectra in Fig. 3a and 3c, it seems that both scenarios reveal similar spectral evolution, however, they involve distinct physics and thus lead to different consequences. Firstly, only the selected mode is damped so that most cavity photons are unaffected in regime II. For example, $Q$ of TE mode is reduced by 25% at $t_r = 68.7$ ps by annihilating less than 5% of the cavity photons in this regime, while the similar modulation of TE mode annihilates more than 15% of the photons (by modulating all the modes of the cavity uniformly) in regime I. Approximate 66% less cavity photon cost is achieved in regime II than conventional approach in regime I for the same level of modulation. Secondly, by setting $t_r \ll 0$ in regime I and with the nanosecond relaxation of photocarriers, the THz probe experiences a quasi-steady process with a fixed carrier density in the entire lifetime of the cavity photons; while in regime II, the 14 ps rising dynamics of temporal boundary suddenly merges into the cavity, and thus results in the perturbation of frequency spectrum as well as the breakdown of Lorentzian lineshape.

**Temporal two-dipole model.** We numerically interpret the dynamic polarization features of the silicon metamaterial. The Drude model in Eq. 1 introduces not only imaginary part of time-varying permittivity of the silicon cavities but also the real part[25], which involve $Q$ factor decline of natural cavity modes as well as generation of new frequency components as a result of the sudden perturbation. The cavity is modelled as a temporal two-dipole emitter: one with the natural resonant polarization of the cavity as a harmonic oscillator subject to a temporal loss boundary (associated to imaginary permittivity) and the other generates harmonics with a frequency-shift polarization after the injection of boundary (associated to real permittivity)[6, 16, 25]:

$$P_1(t) = e^{-i(\omega_1 - i\tau_1)t} \cdot (H(t) - H(t - t_p)) , \tag{3}$$

$$P_2(t) = e^{-i(\omega_1 - i\tau_1)t_p} \cdot e^{-i(\omega_2 - i\tau_2)(t - t_p)} \cdot H(t - t_p) , \tag{4}$$

where $H(t)$ is the Heaviside function in an ideal scenario with a negligible accumulation time relative to the cycle of oscillations, but an exponential function is applied to account for the 14 ps accumulation time of photocarriers in the real model (see supplementary materials). Here $t_0$ is set as the origin of time coordinate, and thus $t_p = t_r$. The total polarization ($P_{tot}(t)$) is the superposition of the two polarizations ($P_{tot}(t) = P_1(t) + P_2(t)$) where the first one describes the cavity mode (at angular frequency $\omega_1$ with damping rate $\tau_1$) radiating at $t = 0$ subject to temporal boundary injected at $t = t_p$, and the second one represents the perturbation induced polarization with a shifted frequency radiating at $t = t_p$ ($\omega_2$ with damping rate $\tau_2$). The first part of $P_2(t)$ is a scaling factor to match the amplitude between the two polarizations at $t_p$. Experimental spectra are well-reproduced by this model in a series of delay times as shown in Fig. 4b. In the time window when TM mode is unperturbed, transmission amplitude difference ($\Delta T = T_0 - T_p$) is merely contributed by the change of TE mode so that it can be well-described by the harmonic perturbation model with fixed constants $\omega_1 = 2\pi \cdot 0.581$ THz, and $\tau_1 = 2\pi \cdot 0.0085$ THz. The frequency-shifted component is fixed at $\omega_2 = 2\pi \cdot 0.590$ THz by fitting the data that would be variable and determined by the carrier density; and $\tau_2$ is the only free parameter in the fitting process[25] (see supplementary materials). According to fitting results of the model, the major contribution of the spectral modulation is from the time-varying imaginary part that causes the linewidth broadening due to the truncation of cavity ringing and the frequency ripples due to the sudden termination of temporal oscillation. It is noted that the frequency ripples are the manifestation of a time-domain filter acting as an apodization function whose frequency period is inversely proportional to $t_p$. The sudden injection of free carriers also induces the change of cavity refractive index as estimated of $\Delta n \approx -0.058$ (with carrier density $\Delta N \approx 1 \times 10^{14}$ cm$^{-3}$) so that a blue-shifted dipole $P_2(t)$ emerges after $t_p$ as a result of the temporal boundary in analogy to changing the tone of guitar by adjusting string length after plucked (see supplementary materials).

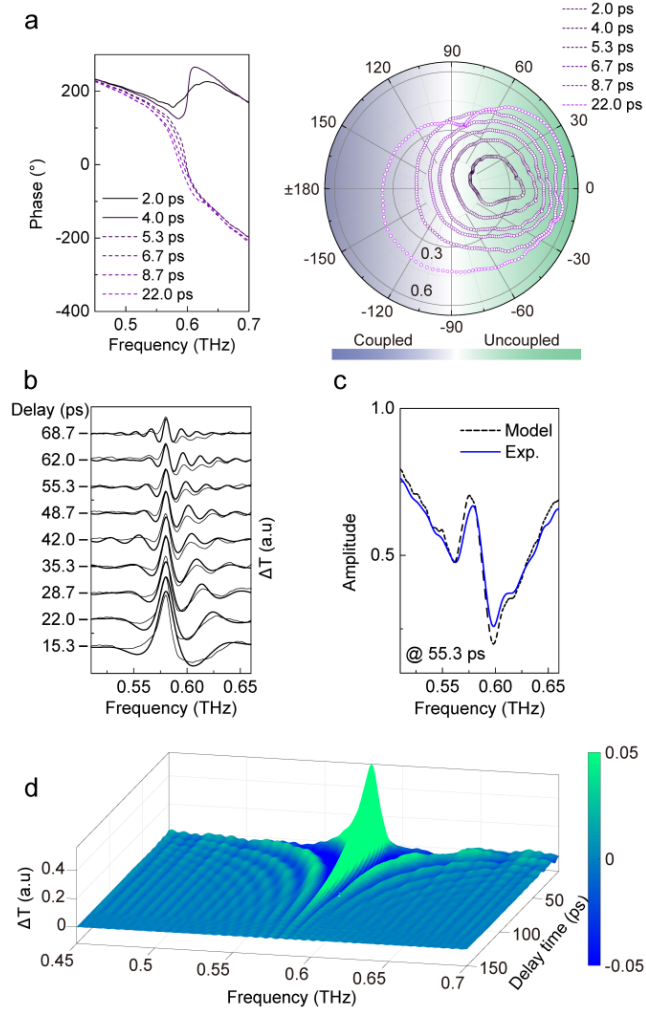

**Figure 4. Transition of coupled cavity and temporal two-dipole model. a.** Phase spectra and the parametric polar plot of transmission for the cavity at various delays indicating the transition of coupled regime. **b.** Comparison of experimental and calculated spectra with the temporal two-dipole model. Most of the spectral features are well-reproduced in the calculation. **c**. Reproduce the experimental transmission spectrum by applying the model to the experimental profile at steady state with $t_p = 55.3$ ps. **d.** Full modulation spectra at various delay times by applying the model to the temporal profile of sample at steady state. Color bar was preset within a small range for a larger color contrast.

The model is directly applied to the experimental steady *temporal* profile (Fig. 2c, and see supplementary materials for temporal profile with pump) in order to numerically reproduce the spectral dynamics. Satisfactory match between model and experimental spectra is obtained using the same preset constants as above in the model with major spectral features well-reproduced (Fig. 4c). Noted that the experimental excitation pulse involves a broadband frequency range (0.1 – 1.4 THz, Fig. 2c), and would excite high-order cavity modes at other frequencies that are not considered in the model. We finally reproduce the full evolution of amplitude transmission difference by tuning the delay $t_\mathrm{p}$ using the model (Fig. 4d). In addition to the clear $t_\mathrm{p}$ dependent ripples in the side bands of the cavity mode, they reveal slightly asymmetric ripple pattern relative to the central frequency of TE mode because of the frequency-shifted polarization $P_2(t)$ (see supplementary materials).

**Discussion**

Although the temporal loss boundary was demonstrated in the terahertz regime, the concept is highly scalable, and would be promising for LIDAR applications.[20] The asynchronized modulation of $Q$-contrasting two-mode cavity can be extended to multimode scenarios with distinct $Q$ values, and a larger contrast would increase the temporal resolution to distinguish the modes. Compared with conventional physical boundary in real space to separate degenerate modes, the temporal boundary approach has no symmetry constraint, which is a transient and reconfigurable process and would especially benefit ultrafast applications. The temporal boundary is not only useful to reduce $Q$, but can also improve $Q$ factor with a smart design of the cavity via storing and regenerating the oscillations. Photon memory would be possible with a controllable store and release time that could break the time-bandwidth limitations via sudden change of cavity properties.[14, 26, 27] Squeezed quantum state could be realized via non-*adiabatic* changes provided that the dynamics is faster than the oscillation cycles.[16] Removal of temporal boundary before the cavity photons are completely damped would be feasible to reach a more flexible engineering of photon dynamics. Such a subcycle control could probably be realized in terahertz regime by applying Kerr nonlinear effect[6, 17, 28] or dynamics of hot carrier thermalization[29] that have been shown to possess sub-picosecond timescale. The free carrier injection or extraction is also feasible by electric pump via electro-optic modulation by integrating PIN diodes[3]. Although we mainly focus on the imaginary part of permittivity, the correlated temporal real part is also involved in the process that contributes to tuning the tone of cavity photons. The temporal boundary induced frequency shift is a linear process and not subject to the phase matching condition, it would therefore be a convenient approach to generate new frequencies whose efficiency can be enhanced by improving cavity $Q$ or working at epsilon near zero wavelength[2,6].

In summary, we have shown the concept of temporal loss boundary as an extra dimension to manipulate the dynamics of cavity photons. The temporal boundary is demonstrated with extended mode control capability than conventional approach. We interpreted the process with a temporal two-dipole mode by merging the sudden perturbation into cavity dynamics that reproduces the experimental spectral features. The temporal dimension introduces an additional degree of freedom to tune the cavity at an ultrafast timescale and would benefit optical communications for adjustable time-bandwidth ratio. The concept may find applications in design of photonic memory and quantum hyperentanglement. The linear tunability of cavity frequency induced by the temporal boundary is not limited by the requirement of the phase matching condition, and would be of special interest in terahertz regime, where nonlinear frequency mixing is challenging due to scarcity of high intensity sources.

**Methods.**

**FDTD simulations.** Data in Fig. 1 was simulated using the finite difference time domain (FDTD) module with periodic boundary conditions for a unit cell. Dimensions of the unit cell are $p = 250$ μm, $d = 240$ μm, and $h = 80$ μm with silicon cylinder sitting on quartz substrate. Lossless materials were set at steady state for silicon and quartz with refractive indices of 3.42 and 2.00, respectively. For simulations subject to temporal boundaries, variation of conductivity in silicon thin film is modelled to account for the dynamic photocarrier density whose thickness is determined by skin depth of pump light. For simplicity, spatially uniform conductivity of the top 10 μm-thick silicon sheet was set to compute the terahertz cavity dynamics in both regime I and regime II. In regime I, a time-invariant conductivity ($\sigma = 100$ S/m for data in Fig. 1d) was defined. In regime II, a time-varying conductivity following a step function $\sigma(t) = \begin{cases} 0 \\ \dfrac{1000}{14}t \text{ S/m} \\ 1000 \end{cases}$ for $\begin{cases} t < t_p \\ t_p \leq t \leq t_p + 14 \\ t > t_p + 14 \end{cases}$ was imported to material property library of the 10 μm-thick silicon sheet. $t_p$ was set at 120 ps for data in Fig. 1d. Electric and magnetic field distributions were extracted from the passive modes at center frequencies with lossless materials.

**Gabor transform.** Gabor transform was used to expand the temporal profile with an extra dimension of frequency so that we can visualize the temporal distribution of frequency components. The transform of the temporal profile $x$(t) is defined by the formula:

$$G_x(\tau,\omega) = \int_{-\infty}^{\infty} x(t) \cdot e^{-\pi(t-\tau)^2} \cdot e^{-j\omega t} dt . \tag{5}$$

In the process of Gabor transform, we perform Fourier transform for partial temporal profile ($x$(t)) that is chosen by the time window defined by Gaussian function. The window should be carefully chosen according the oscillation period of temporal profile for the tradeoff between spectral information and temporal resolution.

**OPTP experiments and dynamics measurements.** Home-built optical pump THz probe (OPTP) spectroscopy setup was used for time-resolved transmission measurements of samples. A near-infrared laser beam (Ti: sapphire laser, 100 fs, 6 mJ per pulse at 800 nm with 1 kHz repetition rate) was coherently split into three parts: external pump source, pump beam for terahertz generation, and detection beam for electro-optical (EO) sampling. All the three beams were synchronized for time-resolved measurements. Transmission spectra without pump were scanned for 200 picoseconds in order to capture all the cavity oscillations (repeated for 10 times and average, but still limited by signal to noise ratio). All the frequency domain spectra were normalized to the reference (a bare substrate) after Fourier transform.

In temporal boundary experiments, height and delay mainly refer to properties of pump pulses. Heigh of boundary is determined by pump fluence, which was adjusted by tuning the variable neutral density filter inserted in the pump beam path as shown in Fig. 5. The relative delay refers to the time difference between terahertz pulses and pump pulses arriving at sample. Time difference was tuned by accurately adjusting the position of retroreflector on the delay stage with a sub-picosecond resolution.

In pump-probe experiments, pump beam diameter (~ 10 mm) is larger than that of the probe terahertz beam so that the probe area is uniformly pumped. By controlling the relative time of terahertz pulse and pump pulse via a delay stage, terahertz time-domain transmission dynamics can be captured by EO sampling. The photocarrier dynamics was measured by monitoring the terahertz peak electric field evolution of time-domain signals (at 2.5 ps, see Fig. 2a) by scanning the arrival time of pump pulses relative to the terahertz pulses. Once the pump pulses arrive at the sample, terahertz peak electric field will reveal perturbations due to injection of photocarriers. The peak electric field evolution was mapped by scanning the pump delay which is proportional to photocarrier density. The longest scanning time is ~3 ns in the measurement of relaxation dynamics of photocarriers due to the limited traveling range of delay stage.

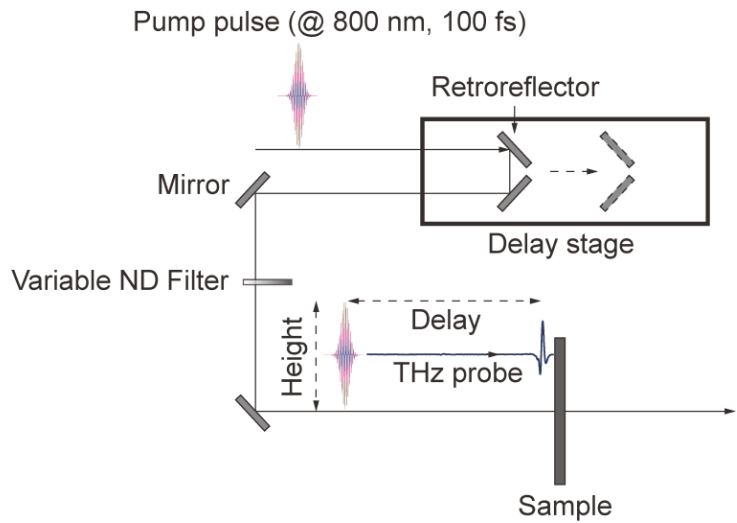

**Figure 5.** Schematic illustration of experimental setup showing how boundary height and delay are adjusted in OPTP system.

**Data availability**

The data that support the findings of this study are available from the corresponding author upon reasonable request.

**Acknowledgements**

This work was supported by the National Natural Science Foundation of China (Award No.: 62175099), and startup funding of Southern University of Science and Technology (SUSTech, Award No.: Y01236147 and Y01236247). The authors acknowledge Singapore Ministry of Education (MOE), MOE2016-T3-1-006 and the National Research Foundation Singapore (Award No.: NRF-CRP23-2019-0005).

**Author contributions**

L.C. conceived ideas and initiated the research. L.C. performed experiments, fabrications, simulations, and analysis. L.C. and R.S. discussed and wrote the manuscript with inputs from J.H. and W.Z. All authors read and commented on the manuscript.

**Competing interests**

The authors declare no competing interests.

# Reference

1. Gunter G., *et al.* Sub-cycle switch-on of ultrastrong light-matter interaction. *Nature* **458**, 178-181 (2009).
2. Tanabe T., Notomi M., Taniyama H., Kuramochi E. Dynamic release of trapped light from an ultrahigh-Q nanocavity via adiabatic frequency tuning. *Phys. Rev. Lett.* **102**, 043907 (2009).
3. Preble S. F., Xu Q., Lipson M. Changing the colour of light in a silicon resonator. *Nat. Photon.* **1**, 293-296 (2007).
4. Hsu C. W., *et al.* Observation of trapped light within the radiation continuum. *Nature* **499**, 188-191 (2013).
5. Akahane Y., Asano T., Song B.-S., Noda S. High-Q photonic nanocavity in a two-dimensional photonic crystal. *Nature* **425**, 944-947 (2003).
6. Zhou Y., *et al.* Broadband frequency translation through time refraction in an epsilon-near-zero material. *Nat. Commun.* **11**, 2180 (2020).
7. Shaltout A. M., Shalaev V. M., Brongersma M. L. Spatiotemporal light control with active metasurfaces. *Science* **364**, (2019).
8. Shaltout A., Kildishev A., Shalaev V. Time-varying metasurfaces and Lorentz non-reciprocity. *Opt. Mater. Express* **5**, 2459 (2015).
9. Guo X., Ding Y., Duan Y., Ni X. Nonreciprocal metasurface with space-time phase modulation. *Light Sci. & Appl.* **8**, 123 (2019).
10. Shao L., *et al.* Non-reciprocal transmission of microwave acoustic waves in nonlinear parity–time symmetric resonators. *Nat. Electron.* **3**, 267-272 (2020).
11. Shaltout A. M., *et al.* Spatiotemporal light control with frequency-gradient metasurfaces. *Science* **365**, 374-377 (2019).
12. Halbhuber M., *et al.* Non-adiabatic stripping of a cavity field from electrons in the deep-strong coupling regime. *Nat. Photon.* **14**, 675-679 (2020).
13. Cong L., Srivastava Y. K., Zhang H., Zhang X., Han J., Singh R. All-optical active THz metasurfaces for ultrafast polarization switching and dynamic beam splitting. *Light Sci. & Appl.* **7**, 28 (2018).
14. Tanaka Y., Upham J., Nagashima T., Sugiya T., Asano T., Noda S. Dynamic control of the Q factor in a photonic crystal nanocavity. *Nat. Mater.* **6**, 862-865 (2007).
15. Lee K., *et al.* Linear frequency conversion via sudden merging of meta-atoms in time-variant metasurfaces. *Nat. Photon.* **12**, 765-773 (2018).
16. McCutcheon M. W., Pattantyus-Abraham A. G., Rieger G. W., Young J. F. Emission spectrum of electromagnetic energy stored in a dynamically perturbed optical microcavity. *Opt. Express* **15**, 11472-11480 (2007).
17. Shcherbakov M. R., Werner K., Fan Z., Talisa N., Chowdhury E., Shvets G. Photon acceleration and tunable broadband harmonics generation in nonlinear time-dependent metasurfaces. *Nat. Commun.* **10**, 1345 (2019).
18. Soref R., Bennett B. *Kramers-Kronig Analysis Of Electro-Optical Switching In Silicon*. SPIE (1987).
19. Cong L., Srivastava Y. K., Solanki A., Sum T. C., Singh R. Perovskite as a Platform for Active Flexible Metaphotonic Devices. *ACS Photonics* **4**, 1595-1601 (2017).

20. Cong L., Singh R. Spatiotemporal Dielectric Metasurfaces for Unidirectional Propagation and Reconfigurable Steering of Terahertz Beams. *Adv. Mater.* **32**, 2001418 (2020).
21. Nagatsuma T., Ducournau G., Renaud C. C. Advances in terahertz communications accelerated by photonics. *Nat. Photon.* **10**, 371-379 (2016).
22. Sengupta K., Nagatsuma T., Mittleman D. M. Terahertz integrated electronic and hybrid electronic–photonic systems. *Nat. Electron.* **1**, 622-635 (2018).
23. Lambert N. J., Rueda A., Sedlmeir F., Schwefel H. G. L. Coherent Conversion Between Microwave and Optical Photons—An Overview of Physical Implementations. *Adv. Quantum Technol.* **3**, 1900077 (2020).
24. Rahimzadegan A.*, et al.* Disorder-Induced Phase Transitions in the Transmission of Dielectric Metasurfaces. *Phys. Rev. Lett.* **122**, 015702 (2019).
25. Soref R., Bennett B. Electrooptical effects in silicon. *IEEE J. Quantum Electron.* **23**, 123-129 (1987).
26. Xu Q., Dong P., Lipson M. Breaking the delay-bandwidth limit in a photonic structure. *Nat. Phy.* **3**, 406-410 (2007).
27. Yanik M. F., Fan S. Stopping light all optically. *Phys Rev Lett* **92**, 083901 (2004).
28. Grinblat G., Berté R., Nielsen M. P., Li Y., Oulton R. F., Maier S. A. Sub-20 fs All-Optical Switching in a Single Au-Clad Si Nanodisk. *Nano Lett.* **18**, 7896-7900 (2018).
29. Yang Y.*, et al.* Femtosecond optical polarization switching using a cadmium oxide-based perfect absorber. *Nat. Photon.* **11**, 390-395 (2017).